\documentclass[aps,preprint,amsmath,amssymb,superscriptaddress,nofootinbib]{revtex4}
\usepackage{multirow}
\usepackage{graphicx}
\begin{document}

\title{Model independent study for the anomalous $W^+W^-\gamma$ couplings at the future lepton-hadron colliders}

\author{S. Spor}
\email[]{serdar.spor@beun.edu.tr}
\affiliation{Department of Medical Imaging Techniques, Zonguldak B\"{u}lent Ecevit University, 67100, Zonguldak, Turkey.}

\author{A. A. Billur}
\email[]{abillur@cumhuriyet.edu.tr} 
\affiliation{Department of Physics, Sivas Cumhuriyet University, 58140, Sivas, Turkey.} 

\author{M. K{\"o}ksal}
\email[]{mkoksal@cumhuriyet.edu.tr} 
\affiliation{Department of Optical Engineering, Sivas Cumhuriyet University, 58140, Sivas, Turkey.}

\begin{abstract}
The triple gauge couplings are completely defined by the non-Abelian gauge nature of the Standard Model, precision measurements of these couplings at the present and future colliders in this way provide a substantial opportunity to test the gauge structure of the Standard Model. Also, measurements of these couplings are sensitive to new physics beyond the Standard Model. In this context, these couplings can be described by an effective Lagrangian. Here, we have studied the potential of the future $\mu p$ colliders on the anomalous $WW\gamma$ interactions via the subprocess $\gamma^* p\,\rightarrow\,W^-\nu_\mu$. This subprocess has been generated through the main process $\mu p\,\rightarrow\,\mu\gamma^* p\,\rightarrow\,W^-\nu_\mu p$ at the LHC-$\mu p$, the FCC-$\mu p$ and the SPPC-$\mu p$. For these reasons, the total cross sections have been obtained as a function of the anomalous $WW\gamma$ couplings. Also, we have been calculated the best constraints on $c_{WWW}$, $c_{W}$ and $c_{B}$ parameters that define the anomalous $WW\gamma$ interactions.
\end{abstract}


\maketitle

\section{Introduction}

The Standard Model (SM) is a successful theory at defining the particle physics phenomena in reachable energy limits of current collider experiments. However, the SM needs to be extended to clarify some problems non-zero neutrino masses, the strong CP problem and matter - antimatter asymmetry in the universe. Therefore, there is a great motivation for new physics research in theoretical and experimental physicists. 

$W^+W^-\gamma$ gauge boson interactions arise from the $SU(2)\times U(1)$ gauge symmetry of the SM. This triple gauge boson coupling is used to test the SM which has been successful in explaining the electroweak theory. Any deviations from the SM values of $W^+W^-\gamma$ coupling indicates the existence of new physics beyond the SM. These discrepancies increase the importance of the anomalous Triple Gauge Boson Couplings (aTGC) of the $W^\pm$ boson such as $W^+W^-\gamma$ and $W^+W^-Z$ in new physics studies and the aTGC have been widely studied in the literature \cite{Baur:1988tfa,Hagiwara:1987xdj,Hagiwara:1992ghg,Wiest:1995yjk,Gintner:1995kkd,Ambrosanio:1992lkj,Sahin:2011dfg,Kepka:2008vmn,Arı:2016aac,Atag:2001ata,Koksal:2019opr,Rodriguez:2019erv,Billur:2019ghy,Sahin:2017uot,Bian:2016wer,Choudhury:1997xsa,Choudhury:1997mep,Li:2018tyb,Kumar:2015ghe,Falkowski:2015ghe,Bhatia:2019gso,Etesami:2016eto,Cakir:2014mjp,Sahin:2009tbz,Atag:2001mai,Bian:2015ylk}. The new physics effects on the anomalous $W^+W^-\gamma$ coupling are investigated in a model independent way via an effective Lagrangian method. Such a method is parameterized by high-dimensional operators which induce the aTGC that modify the interactions between the electroweak gauge bosons. The anomalous $W^+W^-\gamma$ coupling consists of the Lagrangian as dimension-six operators that are made out of SM fields suppressed by the new physics scale $\Lambda$. The effective Lagrangian can be written as:

\begin{eqnarray}
\label{eq.1} 
{\cal L}_{eft}={\cal L}_{SM}+\sum_i\frac{C_i^{(6)}}{\Lambda^2}{\cal O}_i^{(6)}+h.c.
\end{eqnarray}

where ${\cal L}_{SM}$ is renormalize SM Lagrangian, ${C_i^{(6)}}$ are the coupling of ${\cal O}_i^{(6)}$ dimension-six operators and $\Lambda$ is new physics scale. The largest new physics contribution come from dimension-six operators. The effective Lagrangian is written by:

\begin{eqnarray}
\label{eq.2} 
{\cal L}_{eft}=\frac{1}{\Lambda^2}\left[C_W{\cal O}_W+C_B{\cal O}_B+C_{WWW}{\cal O}_{WWW}+h.c.\right]
\end{eqnarray}

where three $C$ and $P$ conserving dimension-six operators are given by:

\begin{eqnarray}
\label{eq.3} 
{\cal O}_{WWW}=Tr\left[W_{\mu\nu}W^{\nu\rho}W_\rho^\mu\right]
\end{eqnarray}
\begin{eqnarray}
\label{eq.4} 
{\cal O}_{W}=\left(D_\mu\Phi\right)^\dagger W^{\mu\nu}\left(D_\nu\Phi\right)
\end{eqnarray}
\begin{eqnarray}
\label{eq.5} 
{\cal O}_{B}=\left(D_\mu\Phi\right)^\dagger B^{\mu\nu}\left(D_\nu\Phi\right)
\end{eqnarray}

where $\Phi$ is the Higgs doublet field. The covariant derivative $D_\mu$ is as follow:

\begin{eqnarray}
\label{eq.6} 
D_\mu \equiv \partial_\mu\,+\,i\frac{g^\prime}{2}B_\mu\,+\,igW_\mu^i\frac{\tau^i}{2}
\end{eqnarray}

where $\tau^i$ are the $SU(2)_I$ generators with Tr$[\tau^i\tau^j]=2\delta^{ij}$; $i,j=1,2,3$. The field strength tensors of the $SU(2)_I$ and $U(1)_Y$ gauge fields are respectively given by:

\begin{eqnarray}
\label{eq.7} 
W_{\mu\nu}=\frac{i}{2}g\tau^i\left(\partial_\mu W_\nu^i - \partial_\nu W_\mu^i + g\epsilon_{ijk}W_\mu^j W_\nu^k\right)
\end{eqnarray}
\begin{eqnarray}
\label{eq.8} 
B_{\mu\nu}=\frac{i}{2}g^\prime\left(\partial_\mu B_\nu - \partial_\nu B_\mu\right)
\end{eqnarray}

where $g$ and $g^\prime$ are $SU(2)_I$ and $U(1)_Y$ couplings, respectively. The effective Lagrangian for $W^+W^-\gamma$ interaction can be then written as follows \cite{Li:2018tyb}:

\begin{eqnarray}
\label{eq.9} 
{\cal L}_{WW\gamma}&=&ig_{WW\gamma}\Big[{g_1^{\gamma}}\left({W_{\mu\nu}^{+}}{W_{\mu}^{-}}A_{\nu}-{W_{\mu\nu}^{-}}{W_{\mu}^{+}}A_{\nu}\right) \nonumber \\
&+&\kappa_{\gamma}{W_{\mu}^{+}}{W_{\nu}^{-}}A_{\mu\nu}+\frac{\lambda_{\gamma}}{M_W^2}{W_{\mu\nu}^{+}}{W_{\nu\rho}^{-}}A_{\rho\mu} \nonumber \\
&+&ig_4^\gamma {W_{\mu}^{+}}{W_{\nu}^{-}}\left( \partial_\mu A_\nu+\partial_\nu A_\mu\right) \\
&-&ig_5^\gamma \epsilon_{\mu\nu\rho\sigma}\left({W_{\mu}^{+}}\partial_\rho{W_{\nu}^{-}}-\partial_\rho{W_{\mu}^{+}}{W_{\nu}^{-}}\right)A_\sigma \nonumber \\
&+&\tilde{\kappa}_{\gamma}{W_{\mu}^{+}}{W_{\nu}^{-}}\tilde{A}_{\mu\nu}+\frac{\tilde{\lambda}_{\gamma}}{M_W^2}{W_{\lambda\mu}^{+}}{W_{\mu\nu}^{-}}\tilde{A}_{\nu\lambda} \Big] \nonumber 
\end{eqnarray}

where $g_{WW\gamma}=-e$ and $\tilde{A}=\frac{1}{2}\epsilon_{\mu\nu\rho\sigma}A_{\rho\sigma}$. $A^{\mu\nu}=\partial^\mu A^\nu - \partial^\nu A^\mu$ is the field strength tensor for photon. $g_1^{\gamma}$, $\kappa_{\gamma}$ and $\lambda_{\gamma}$ anomalous parameters at the first three terms of Eq.~(\ref{eq.9}) are both $C$ and $P$ conserving and $g_4^\gamma$, $g_5^\gamma$, $\tilde{\kappa}_{\gamma}$ and $\tilde{\lambda}_{\gamma}$ anomalous parameters at the remaining four terms are $C$ and/or $P$ violating. Electromagnetic gauge invariance requires that ${g_1^{\gamma}}=1$. In the SM, the anomalous coupling parameters are given by $\kappa_{\gamma}=1$ and $\lambda_{\gamma}=0$. There are only two anomalous parameters, $\kappa_{\gamma}$ and $\lambda_{\gamma}$, in the absence of $C$ and/or $P$ violation beyond the SM. If the anomalous coupling parameters in the effective Lagrangian are reconsidered as six dimensional operators, the desired properties of the effective field theory remain unchanged. The parameters can be reframed and transformed into $c_{WWW}$, $c_{W}$ and $c_{B}$ \cite{Degrande:2013rry}. Thus, the effective field theory approach allow the following coefficients to be expressed:

\begin{eqnarray}
\label{eq.10} 
{\kappa_\gamma}=1+\left(c_W+c_B\right)\frac{m_W^2}{2\Lambda^2}\,,
\end{eqnarray}
\begin{eqnarray}
\label{eq.11} 
{\lambda_\gamma}=c_{WWW}\frac{3g^2m_W^2}{2\Lambda^2}\,.
\end{eqnarray}

Here, $c_{WWW}$, $c_{W}$ and $c_{B}$ parameters determine new physics contributions. In the SM, the anomalous coupling parameters are given by $c_{WWW}=0$, $c_{W}=0$ and $c_{B}=0$.

The anomalous $W^+W^-\gamma$ couplings have been studied on the parameters of $\kappa_{\gamma}$ and $\lambda_{\gamma}$ at the LEP \cite{Schael:2013msh}, the Tevatron \cite{Aaltonen:2009skw,Abazov:2012mwc} and the LHC experiments \cite{Aaboud:2017les,Sirunyan:2017txe,Sirunyan:2019umc}. 

\section{Muon-proton colliders}

Studies of accelerator physics that deal with various types of collisions make significant contributions to new physics research in particle physics. Hadron colliders are called discovery machines that have the highest center-of-mass values, whereas the lepton colliders and the lepton-hadron colliders are known as precision machines have generally the lower ones. The LHC is a high potential machine for discovering new particles and interactions. However, precise measurements are difficult to perform in the LHC, as the large number of jets scattered after the collision of the proton beams causes the backgrounds or noises complicated to detect the sought signals. The LHC, the most powerful and largest circular $pp$ collider ever built, will be developed gradually with the developing accelerator technology. Nevertheless, the search for new physics at beyond the SM makes the lepton-hadron colliders as an important potential candidate in the future of particle physics. In this post-LHC process, it is planned that the LHC will be first transformed into the Large Hadron electron Collider (LHeC) having an electron ring to be tangentially constructed to the main tunnel of the LHC and after the completion of the LHeC programme, the LHC-$\mu p$ will be operated as new lepton-hadron colliders by replacing the electron ring with the muon ring \cite{Alici:2019jhj}. The Electron-Ion Collider (EIC) is planned a type of particle accelerator where spin-polarized beams of electrons and ions collide to examine the properties of nuclear matter in detail through deep inelastic scattering \cite{Accardi:2016rty}. In January 2020, the U.S. Department of Energy (DOE) approved Brookhaven National Laboratory in New York as the site for building the EIC. On the other hand, the Future Circular Collider (FCC) is considered a circular collider at CERN for the post-LHC era. In the FCC project, the first step involves the design of a future $e^-e^+$ collider with center-of-mass energy of 90-365 GeV, while the addition of $pp$, $ep$, $\mu\mu$ and $\mu p$ colliders is also contemplated \cite{Abada:2019erv}. The FCC-hh is planned as a future $pp$ collider with $\sqrt{s}=100$ TeV \cite{Abada:2019hjp}. Construction of muon collider by adding a muon ring tangential to the FCC will enable to use of high proton energy. Thus, it provides an opportunity to investigate lepton-hadron collisions at high center-of-mass energy with $\mu p$ and $\mu A$ colliders \cite{Caliskan:2017erw}. Chinese scientists have designed a $pp$ collider, namely Super Proton Proton Collider (SPPC), with a center-of-mass energy of 70 TeV in parallel to the FCC project. Before the SPPC collider, the Circular Electron Positron Collider (CEPC) that is an future $e^-e^+$ collider is designed as the first stage using the same tunnel. In the CEPC/SPPC project, $\mu\mu$, $\mu p$ and $ep$ collisions are also performed in the later years, such as the FCC project \cite{CEPC:2018wev}. Below we list past and future energy frontier colliders for two time periods:

Before the LHC ($<$2010): Tevatron \cite{Fermilab:1984ger}, SLC/LEP \cite{SLAC:1980opr,LEP:1983tty}, HERA \cite{HERA:1981fmq},

LHC era and beyond the LHC ($>$2010): LHC \cite{Evans:2008tyu}, ILC \cite{ILC:2013ywe}, $\mu$C ($\mu^-\mu^+$) \cite{Delahaye:2013ujk}, CEPC \cite{CEPC:2018wev}, LHeC \cite{Fernandez:2012xcv}, FCC-ee \cite{Bicer:2014ytr}, CLIC \cite{Linssen:2012era}, FCC-hh \cite{FCC:2017ghw}, SPPC \cite{SPPC:2015lhb}, FCC-he \cite{Oliver:2017qps}. 

$\mu p$ colliders make it possible to use $\gamma^*\gamma^*$, $\gamma^*\mu$ and $\gamma^*p$ interactions possible to study the new physics beyond the SM. The emitted photons from the incoming protons scattering at very small angles from the beam pipe. Thus, since these photons have very low virtuality, they are almost-real. The Equivalent Photon Approximation (EPA) \cite{Budnev:1975gvr,Terazawa:1973hht} is a facility in phenomenological investigations because it permits to obtain cross sections for the process $\gamma^*\mu\,\rightarrow\,X$ approximately via the study of the process $\mu p\,\rightarrow\,\mu^-\gamma^* p\,\rightarrow\,X p$ process where $X$ shows particles obtained in the final state. In addition, these interactions have very clean experimental conditions. 

In this study, we have investigated the anomalous $W^+W^-\gamma$ couplings through the process $\mu p\,\rightarrow\,\mu\gamma^* p\,\rightarrow\,W^-\nu_\mu p$ and calculated the cross sections and constraints on the anomalous $c_{WWW}$, $c_{W}$ and $c_{B}$ coupling parameters in several different channels and colliders.

\section{Cross sections and sensitivity analysis of the process $\mu p\,\rightarrow\,\mu\gamma^* p\,\rightarrow\,W^-\nu_\mu p$ at the LHC, the FCC and the SPPC}

\begin{figure}
\includegraphics[scale=0.3]{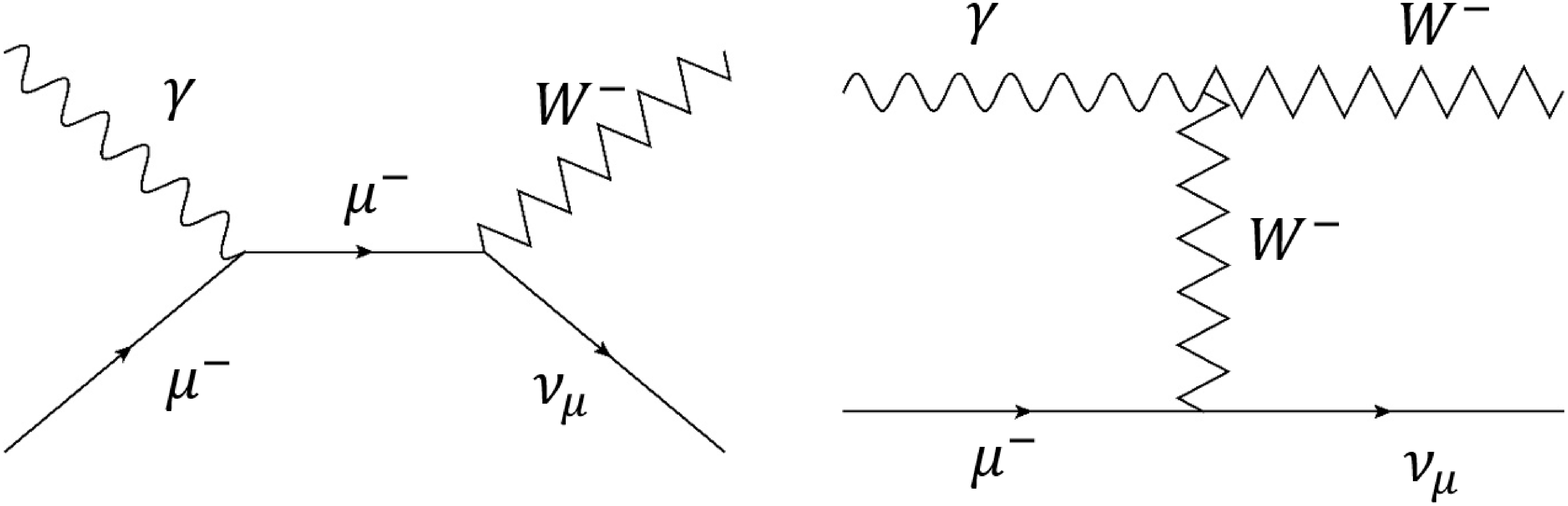}
\caption{Tree-level Feynman diagrams for the subprocess $\mu\gamma^*\,\rightarrow\,W^-\nu_\mu$.}
\label{fig1}
\end{figure}

The main process $\mu p\,\rightarrow\,\mu\gamma^* p\,\rightarrow\,W^-\nu_\mu p$ consists of the subprocess $\mu\gamma^*\,\rightarrow\,W^-\nu_\mu$. The subprocess $\mu\gamma^*\,\rightarrow\,W^-\nu_\mu$ is described by two tree level Feynman diagrams given in Fig.~\ref{fig1}. Only one of these Feynman diagrams has the anomalous $W^+W^-\gamma$ coupling arising from new physics effects. The calculations of cross section in this paper are simulated using {\sc MadGraph5} an aMC@NLO \cite{Alwall:2014cvc}. Here, we use the EWdim6 model file for the operators that examine interactions between the electroweak gauge boson we have described above in dimension-six \cite{Degrande:2013rry}. The used values for muon energy, proton energy, center-of-mass energy and integrated luminosity at the LHC-$\mu p$ \cite{Alici:2019jhj,Ozansoy:2019oaz}, the FCC-$\mu p$ \cite{Acar:2018huo,Caliskan:2017erw,Acar:2017ggs} and the SPPC-$\mu p$ \cite{Canbay:2017ion} are given in Table~\ref{tab1}. 

\begin{figure}
\includegraphics[scale=0.6]{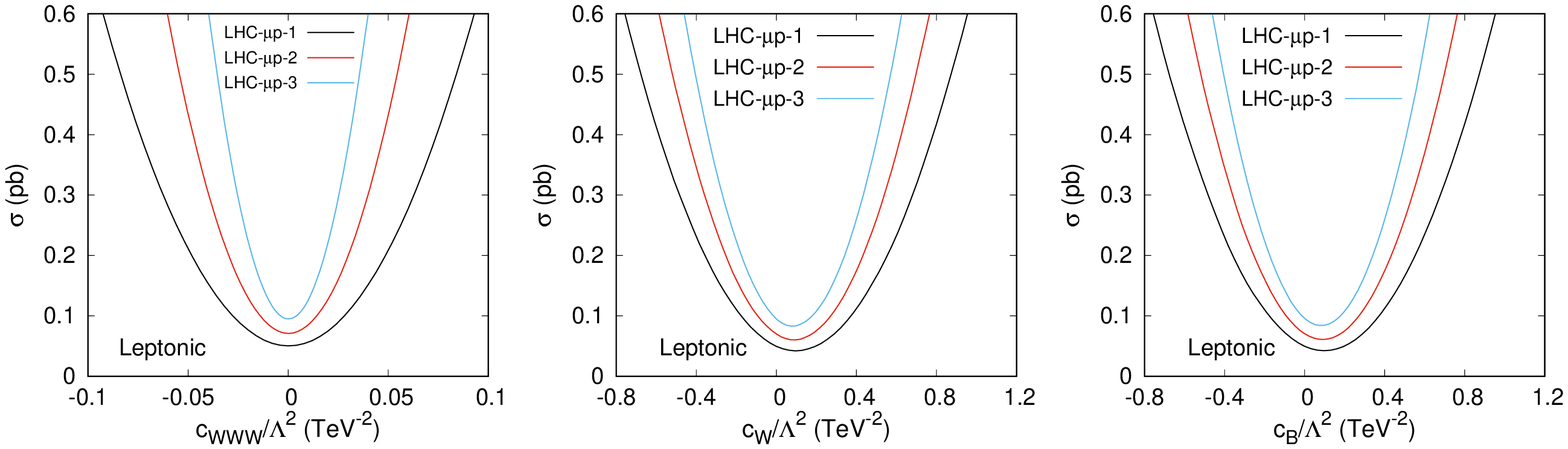}
\caption{The total cross sections of leptonic decay channel of the main process $\mu p\,\rightarrow\,\mu\gamma^* p\,\rightarrow\,W^-\nu_\mu p$ as a function of $c_{WWW}$, $c_{W}$ and $c_{B}$ for muon energies of $E_\mu=0.75,\,1.5,\,3$ TeV at the LHC-$\mu p$.}
\label{fig2}
\end{figure}

\begin{figure}
\includegraphics[scale=0.6]{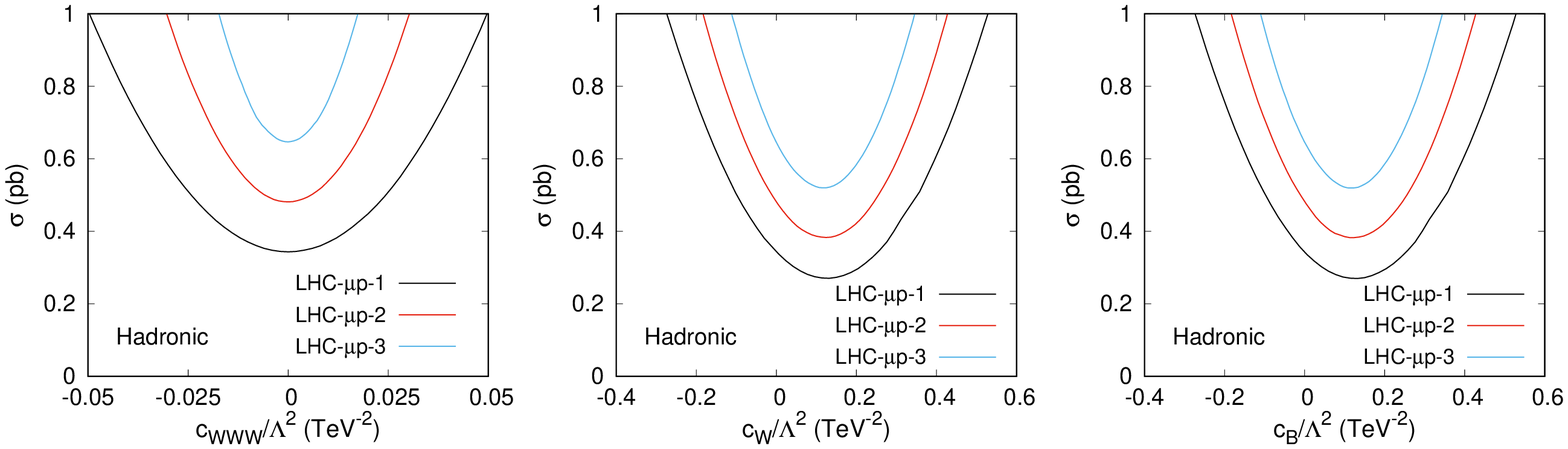}
\caption{Same as in Fig.~\ref{fig2}, but for hadronic decay channel.}
\label{fig3}
\end{figure}

\begin{figure}
\includegraphics[scale=0.6]{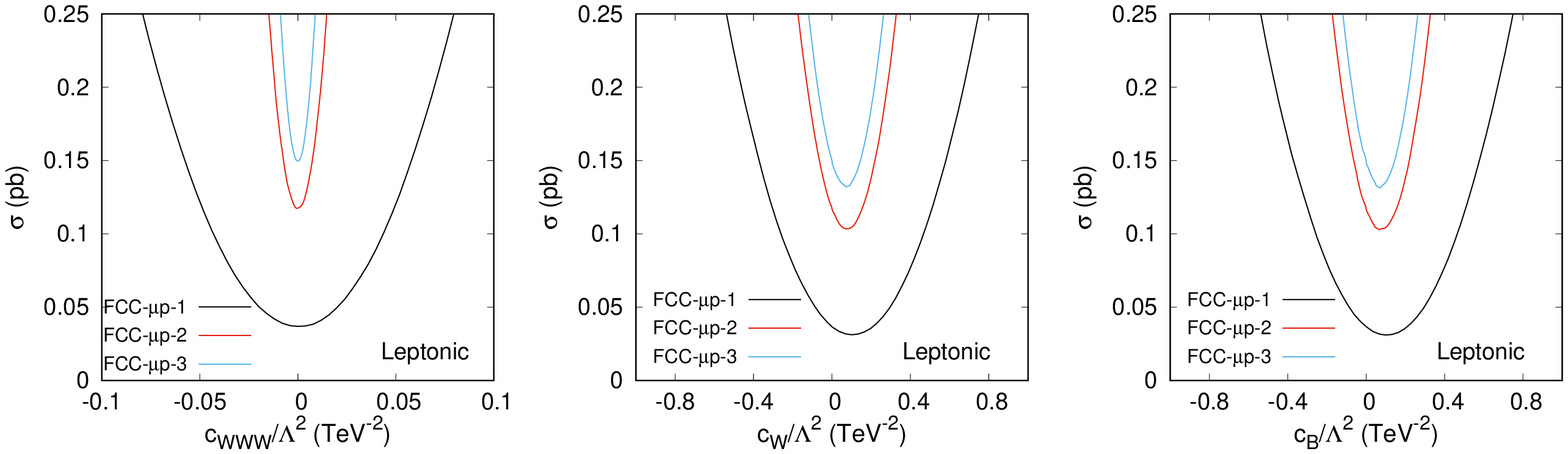}
\caption{The total cross sections of leptonic decay channel of the main process $\mu p\,\rightarrow\,\mu\gamma^* p\,\rightarrow\,W^-\nu_\mu p$ as a function of $c_{WWW}$, $c_{W}$ and $c_{B}$ for muon energies of $E_\mu=0.063,\,0.75,\,1.5$ TeV at the FCC-$\mu p$.}
\label{fig4}
\end{figure}

\begin{figure}
\includegraphics[scale=0.6]{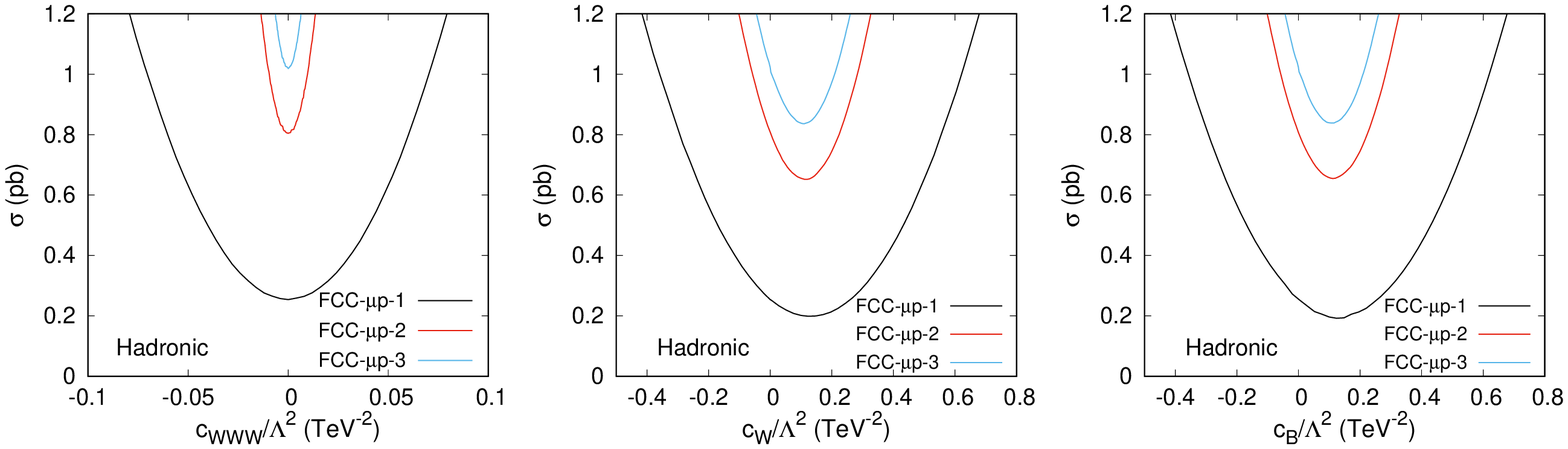}
\caption{Same as in Fig.~\ref{fig4}, but for hadronic decay channel.}
\label{fig5}
\end{figure}

\begin{figure}
\includegraphics[scale=0.6]{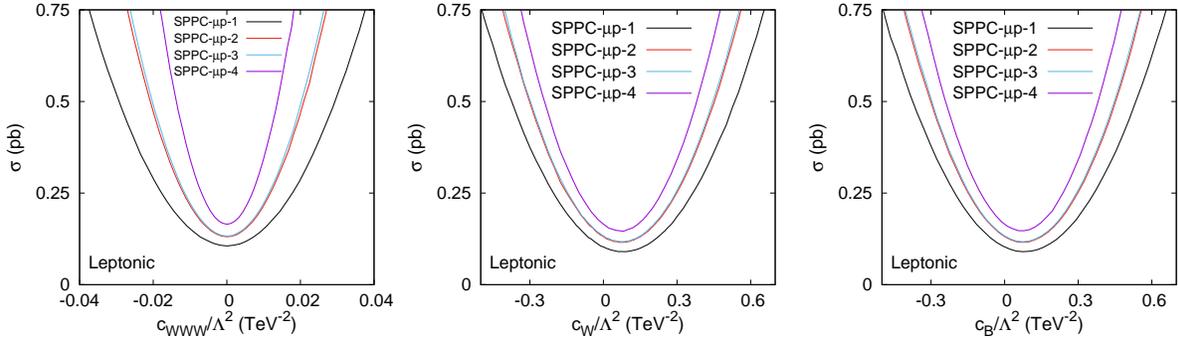}
\caption{The total cross sections of leptonic decay channel of the main process $\mu p\,\rightarrow\,\mu\gamma^* p\,\rightarrow\,W^-\nu_\mu p$ as a function of $c_{WWW}$, $c_{W}$ and $c_{B}$ at the SPPC-$\mu p$.}
\label{fig6}
\end{figure}

\begin{figure}
\includegraphics[scale=0.6]{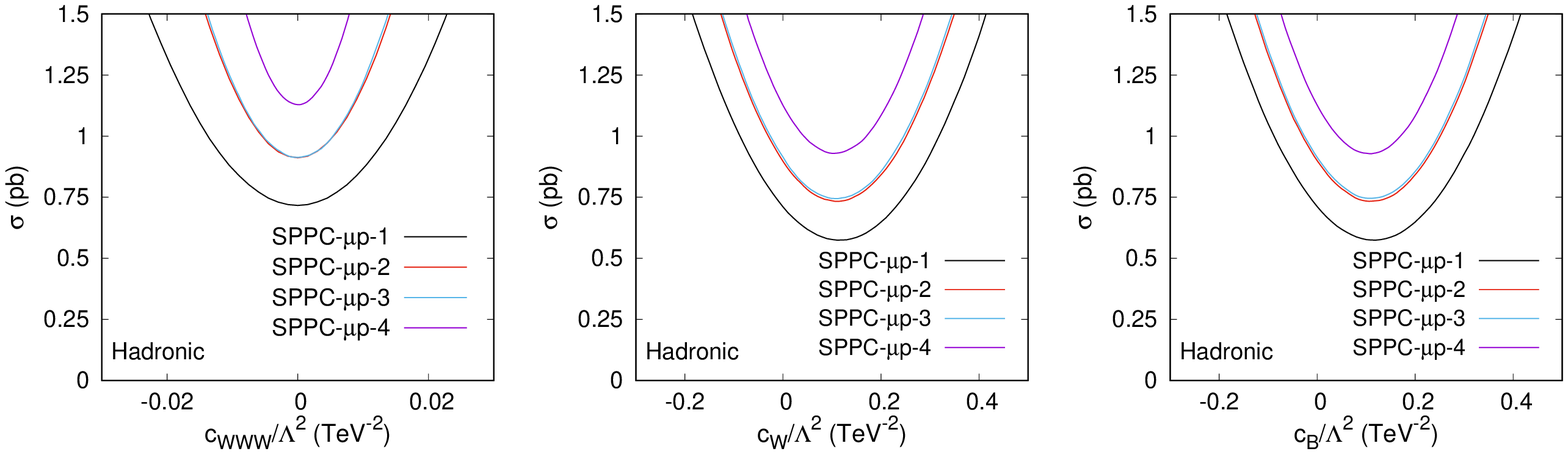}
\caption{Same as in Fig.~\ref{fig6}, but for hadronic decay channel.}
\label{fig7}
\end{figure}

We assume that the $W^-$ bosons produced from the subprocess $\mu\gamma^*\,\rightarrow\,W^-\nu_\mu$ have the leptonic or hadronic decay channels. We consider $W^-\,\rightarrow\,{\ell} \overline{\nu}_{\ell}$ for leptonic decay and $W^-\,\rightarrow\,q \overline{q}$ for hadronic decay, where ${\ell}=e^-,\,\mu^-$ and $q=b,\,c,\,d,\,s,\,u$. The total cross sections of the main process $\mu p\,\rightarrow\,\mu\gamma^* p\,\rightarrow\,W^-\nu_\mu p$ as a function of $c_{WWW}$, $c_{W}$ and $c_{B}$ for leptonic and hadronic decay channels are presented in Figs.~\ref{fig2} and \ref{fig3} at the LHC-$\mu p$ colliders, in Figs.~\ref{fig4} and \ref{fig5} at the FCC-$\mu p$ colliders and in Figs.~\ref{fig6} and \ref{fig7} at the SPPC-$\mu p$ colliders, respectively. When the figures are examined, the values of the total cross section with hadronic decay are higher in all colliders than the ones with leptonic decay and increase as the muon energy increases. The characteristics of the total cross sections as a function of the anomalous $c_{W}$ and $c_{B}$ couplings parameters are similar due to Eq.~(\ref{eq.10}).

With the help of statistical analysis, it is possible to determine deviations from the predictions of the SM in the cross sections caused by new physics contributions. Therefore, $\chi^2$ analysis is performed to obtain the constraints on the anomalous coupling parameters at the 95$\%$ Confidence Level (C.L.). $\chi^2$ function is defined by \cite{Hernandez:2019aaa,Koksal:2019mkn,Billur:2019mkn}:

\begin{eqnarray}
\label{eq.12} 
\chi^2=\left(\frac{\sigma_{SM}-\sigma_{NP}}{\sigma_{SM}\delta}\right)^2\,.
\end{eqnarray}

Here, $\sigma_{SM}$ is the cross section in the SM and $\sigma_{NP}$ is the cross section containing the SM and new physics contributions. $\delta=\frac{1}{\sqrt {N_{SM}}}$ is the statistical error. The number of SM events is presented by $N_{SM}=L\times \sigma_{SM}$, where $L$ is the integrated luminosity.

Moreover, statistical significance (SS) analysis is defined to obtain the constraints on the anomalous coupling parameters at the 99$\%$ C.L. as \cite{Caliskan:2017ope}

\begin{eqnarray}
\label{eq.13} 
SS=\frac{\left|\sigma_{NP}-\sigma_{SM}\right|}{\sqrt{\sigma_{SM}}}\sqrt{L}
\end{eqnarray}

where definitions of $\sigma_{SM}$, $\sigma_{NP}$ and $L$ are the same as in the paragraph above.

The constraints on the anomalous coupling parameters named as ``$\chi^2$ analysis'' and ``SS analysis'' in Tables~\ref{tab2}-\ref{tab3} are determined by $\chi^2\ge3.84$ at the 95$\%$ C.L. on Eq.~(\ref{eq.12}) and $SS\ge5$ at the 99$\%$ C.L. on Eq.~(\ref{eq.13}), respectively. We have presented constraints for the anomalous $c_{WWW}$, $c_{W}$ and $c_{B}$ coupling parameters at the LHC-$\mu p$, the FCC-$\mu p$ and the SPPC-$\mu p$ and analyzed as leptonic and hadronic decay channels in Tables~\ref{tab2}-\ref{tab3}. 

Furthermore, as a result of $\chi^2$ and SS analysis, we have obtained the best constraints on the center-of-mass energies and integrated luminosities values of the LHC-$\mu p$-3, the FCC-$\mu p$-3 and the SPPC-$\mu p$-4 colliders for each of the anomalous coupling parameters and the most sensitive of these colliders is the SPPC-$\mu p$ collider as seen in Tables~\ref{tab2}-\ref{tab3}. The sensitivities on the anomalous $c_{WWW}$ coupling parameter in the hadronic decay channels of the process $\mu p\,\rightarrow\,\mu\gamma^* p\,\rightarrow\,W^-\nu_\mu p$ are about 1.2 times better than the leptonic ones at all of the colliders. However, in general although maximum constraints of the sensitivities on the anomalous $c_{W}$ and $c_{B}$ coupling parameters in the leptonic decay channels are about 1.3 times more sensitive than hadronic ones, minimum constraints of the sensitivities on the anomalous $c_{W}$ and $c_{B}$ coupling parameters in the hadronic decay channels are about 2.7 times more sensitive than leptonic ones. The sensitivities obtained using the SS analysis for the anomalous $c_{WWW}$ coupling parameter in Table~\ref{tab2} is 2 times worse than that using the $\chi^2$ analysis. While in Table~\ref{tab3} the minimum constraints of the obtained sensitivities using SS analysis for the anomalous $c_{W}$ and $c_{B}$ coupling parameters are 2 times worse than that using $\chi^2$ analysis, the maximum constraints are approximately the same order in both analyzes.

\section{Conclusions}

The aTGC is one of the mechanisms that allows the research to test the SM an beyond the SM. We have investigated the anomalous $WW\gamma$ interactions through the main process $\mu p\,\rightarrow\,\mu\gamma^* p\,\rightarrow\,W^-\nu_\mu p$ at the future $\mu p$ colliders that are planned to be built in the future years. The cross sections of the LHC-$\mu p$, the FCC-$\mu p$ and the SPPC-$\mu p$ colliders with different center-of-mass energies and integrated luminosities are calculated in both leptonic and hadronic decay channels as a function of the anomalous $c_{WWW}$, $c_{W}$ and $c_{B}$ coupling parameters. Thus, the contribution of various $\mu p$ colliders to the process is determined in each anomalous coupling parameter. Using the $\chi^2$ and SS analysis, we have revealed the constraints on $c_{WWW}$, $c_{W}$ and $c_{B}$ parameters for different center-of-mass energies and integrated luminosities of the LHC-$\mu p$, the FCC-$\mu p$ and the SPPC-$\mu p$ colliders. Ref.~\cite{Sirunyan:2019umc} has established updated constraints on the anomalous $W^+W^-\gamma$ and $W^+W^-Z$ couplings with the parameters of $c_{WWW}$, $c_{W}$ and $c_{B}$ and the constraints are determined to be $-1.58$ TeV$^{-2}<c_{WWW}/\Lambda^2<1.59$ TeV$^{-2}$, $-2.00$ TeV$^{-2}<c_{W}/\Lambda^2<2.65$ TeV$^{-2}$ and $-8.78$ TeV$^{-2}<c_{B}/\Lambda^2<8.54$ TeV$^{-2}$ by the CMS experiment at the CERN LHC. We have proved that the sensitivities on the anomalous $c_{WWW}$, $c_{W}$ and $c_{B}$ coupling parameters are successful enough for all of the colliders by comparing our sensitivity results with the Ref.~\cite{Sirunyan:2019umc}. If the LHC-$\mu p$-3, the FCC-$\mu p$-3 and the SPPC-$\mu p$-4 collider results in Tables~\ref{tab2}-\ref{tab3} are compared with the sensitivities of the Ref.~\cite{Sirunyan:2019umc}, in both the leptonic and hadronic decay channels at Tables~\ref{tab2}-\ref{tab3} the sensitivities on the anomalous $c_{WWW}$ coupling parameter are about 385, 555 and 1150 times, that of the anomalous $c_{W}$ coupling parameter are about 21, 22 and 25 times and that of the anomalous $c_{B}$ coupling parameter are about 75, 85 and 100 times better than the sensitivities of the Ref.~\cite{Sirunyan:2019umc}, respectively. We have compared the colliders according to their sensitivity and concluded that the collider having the best sensitivity is the SPPC-$\mu p$-4 collider. On the other hand, if the comparison is repeated with the lowest center-of-mass energy the LHC-$\mu p$-1, the FCC-$\mu p$-1 and the SPPC-$\mu p$-1 colliders, it is seen that the sensitivities are still better than the Ref.~\cite{Sirunyan:2019umc} although they decrease. Therefore, we have determined that the LHC-$\mu p$, the FCC-$\mu p$ and the SPPC-$\mu p$ colliders provide new opportunities to investigate the anomalous $WW\gamma$ couplings.

\newpage

\begin{table}
\caption{The used values of muon-proton colliders.}
\label{tab1}
\begin{ruledtabular}
\begin{tabular}{ccccc}
Colliders&$E_\mu$ (TeV)&$E_p$ (TeV)&$\sqrt s$ (TeV)&$L$ ($fb^{-1}$)\\
\hline \hline
LHC-$\mu p$-1&0.75&7.0&4.58&14\\
LHC-$\mu p$-2&1.5&7.0&6.48&23\\
LHC-$\mu p$-3&3.0&7.0&9.16&9\\
\hline
FCC-$\mu p$-1&0.063&50.0&3.50&0.02\\
FCC-$\mu p$-2&0.75&50.0&12.2&5\\
FCC-$\mu p$-3&1.5&50.0&17.3&5\\
\hline
SPCC-$\mu p$-1&0.75&35.6&10.33&5.5\\
SPCC-$\mu p$-2&0.75&68.0&14.28&12.5\\
SPCC-$\mu p$-3&1.5&35.6&14.61&4.9\\
SPCC-$\mu p$-4&1.5&68.0&20.20&42.8\\
\end{tabular}
\end{ruledtabular}
\end{table}

\begin{table}
\caption{95\% C.L. and 99\% C.L. constraints on the anomalous $c_{WWW}$ coupling at the LHC-$\mu p$, the FCC-$\mu p$ and the SPPC-$\mu p$. The used parameters of various colliders are given by Table~\ref{tab1}. The leptonic and hadronic decay channels of the process $\mu p\,\rightarrow\,\mu\gamma^* p\,\rightarrow\,W^-\nu_\mu p$ are considered separately.}
\label{tab2}
\begin{ruledtabular}
\begin{tabular}{cccc}
\multicolumn{2}{c}{} & \multicolumn{2}{c}{$c_{WWW}$} \\
\hline
Colliders & Analysis & Leptonic & Hadronic \\ 
\hline \hline
\multirow{2}{*}{LHC-$\mu p$-1} 
 & $\chi^2$ & [-0.00745; 0.00783] & [-0.00601; 0.00607] \\ 
 & SS & [-0.01202; 0.01240] & [-0.00962; 0.00968] \\ \hline
\multirow{2}{*}{LHC-$\mu p$-2} 
 & $\chi^2$ & [-0.00471; 0.00503] & [-0.00394; 0.00400] \\ 
 & SS & [-0.00762; 0.00793] & [-0.00632; 0.00638] \\ \hline
\multirow{2}{*}{LHC-$\mu p$-3} 
 & $\chi^2$ & [-0.00442; 0.00446] & [-0.00378; 0.00371] \\ 
 & SS & [-0.00707; 0.00711] & [-0.00601; 0.00596] \\
\hline \hline
\multirow{2}{*}{FCC-$\mu p$-1} 
 & $\chi^2$ & [-0.04912; 0.04964] & [-0.03833; 0.03833] \\ 
 & SS & [-0.07861; 0.07902] & [-0.06125; 0.06124] \\ \hline
\multirow{2}{*}{FCC-$\mu p$-2} 
 & $\chi^2$ & [-0.00394; 0.00403] & [-0.00339; 0.00344] \\ 
 & SS & [-0.00632; 0.00641] & [-0.00543; 0.00548] \\ \hline
\multirow{2}{*}{FCC-$\mu p$-3} 
 & $\chi^2$ & [-0.00285; 0.00291] & [-0.00250; 0.00250] \\ 
 & SS & [-0.00457; 0.00463] & [-0.00402; 0.00402] \\
\hline \hline
\multirow{2}{*}{SPPC-$\mu p$-1} 
 & $\chi^2$ & [-0.00431; 0.00437] & [-0.00385; 0.00382] \\ 
 & SS & [-0.00690; 0.00696] & [-0.00614; 0.00612] \\ \hline
\multirow{2}{*}{SPPC-$\mu p$-2} 
 & $\chi^2$ & [-0.00269; 0.00279] & [-0.00236; 0.00239] \\ 
 & SS & [-0.00433; 0.00443] & [-0.00378; 0.00381] \\ \hline
\multirow{2}{*}{SPPC-$\mu p$-3} 
 & $\chi^2$ & [-0.00333; 0.00342] & [-0.00294; 0.00294] \\ 
 & SS & [-0.00535; 0.00544] & [-0.00470; 0.00470] \\ \hline
\multirow{2}{*}{SPPC-$\mu p$-4} 
 & $\chi^2$ & [-0.00146; 0.00148] & [-0.00128; 0.00131] \\ 
 & SS & [-0.00234; 0.00236] & [-0.00205; 0.00208] \\
\end{tabular}
\end{ruledtabular}
\end{table}

\begin{table}
\caption{Same as in Table~\ref{tab2}, but for the anomalous $c_{W}$ and  $c_{B}$ couplings.}
\label{tab3}
\begin{ruledtabular}
\begin{tabular}{cccccc}
\multicolumn{2}{c}{} & \multicolumn{2}{c}{$c_{W}$} & \multicolumn{2}{c}{$c_{B}$} \\
\hline
Colliders & Anal. & Leptonic & Hadronic & Leptonic & Hadronic \\ 
\hline \hline
\multirow{2}{*}{LHC-$\mu p$-1} 
 & $\chi^2$ & [-0.02218; 0.21885] & [-0.00812; 0.26098] & [-0.02239; 0.21612] & [-0.00811; 0.26127]\\ 
 & SS & [-0.05018; 0.24685] & [-0.01983; 0.27274] & [-0.05054; 0.24432] & [-0.01980; 0.27300]\\ \hline
\multirow{2}{*}{LHC-$\mu p$-2} 
 & $\chi^2$ & [-0.01470; 0.19456] & [-0.00543; 0.24969] & [-0.01498; 0.19409] & [-0.00540; 0.24941]\\ 
 & SS & [-0.03412; 0.21398] & [-0.01342; 0.25768] & [-0.03470; 0.21381]  & [-0.01337; 0.25738]\\ \hline
\multirow{2}{*}{LHC-$\mu p$-3} 
 & $\chi^2$ & [-0.01962; 0.18485] & [-0.00742; 0.24118] & [-0.01953; 0.18662] & [-0.00742; 0.24086]\\ 
 & SS & [-0.04420; 0.20945] & [-0.01815; 0.25193] & [-0.04407; 0.21113] & [-0.01813; 0.25162]\\
\hline \hline
\multirow{2}{*}{FCC-$\mu p$-1} 
 & $\chi^2$ & [-0.30422; 0.51306] & [-0.15778; 0.42019] & [-0.30361; 0.51143] & [-0.15805; 0.42031]\\
 & SS & [-0.53481; 0.74486] & [-0.30060; 0.56295] & [-0.53448; 0.74374] & [-0.30089; 0.56280]\\ \hline
\multirow{2}{*}{FCC-$\mu p$-2} 
 & $\chi^2$ & [-0.02291; 0.17827] & [-0.00883; 0.23487] & [-0.02271; 0.17800] & [-0.00892; 0.23476]\\ 
 & SS & [-0.05057; 0.20594] & [-0.02143; 0.24738] & [-0.05021; 0.20546] & [-0.02161; 0.24742]\\ \hline
\multirow{2}{*}{FCC-$\mu p$-3} 
 & $\chi^2$ & [-0.02013; 0.16413] & [-0.00799; 0.22367] & [-0.02023; 0.16319] & [-0.00794; 0.22407]\\ 
 & SS & [-0.04474; 0.18877] & [-0.01944; 0.23511] & [-0.04481; 0.18801] & [-0.01927; 0.23522]\\
\hline \hline
\multirow{2}{*}{SPPC-$\mu p$-1} 
 & $\chi^2$ & [-0.02329; 0.18471] & [-0.00901; 0.24002] & [-0.02298; 0.18514] & [-0.00900; 0.24004]\\ 
 & SS & [-0.05155; 0.21295] & [-0.02182; 0.25280] & [-0.05096; 0.21311] & [-0.02181; 0.25284]\\ \hline
\multirow{2}{*}{SPPC-$\mu p$-2} 
 & $\chi^2$ & [-0.01428; 0.16401] & [-0.00542; 0.22642] & [-0.01434; 0.16459] & [-0.00541; 0.22710]\\ 
 & SS & [-0.03274; 0.18249] & [-0.01335; 0.23441] & [-0.03288; 0.18313] & [-0.01335; 0.23502]\\ \hline
\multirow{2}{*}{SPPC-$\mu p$-3} 
 & $\chi^2$ & [-0.02173; 0.17026] & [-0.00842; 0.22858] & [-0.02163; 0.17095] & [-0.00848; 0.22939]\\
 & SS & [-0.04801; 0.19658] & [-0.02041; 0.24066] & [-0.04786; 0.19720] & [-0.02056; 0.24144]\\ \hline
\multirow{2}{*}{SPPC-$\mu p$-4} 
 & $\chi^2$ & [-0.00717; 0.14632] & [-0.00266; 0.21492] & [-0.00721; 0.14576] & [-0.00265; 0.21495]\\ 
 & SS & [-0.01712; 0.15629] & [-0.00667; 0.21893] & [-0.01721; 0.15581] & [-0.00664; 0.21893]\\
\end{tabular}
\end{ruledtabular}
\end{table}

\end{document}